\documentclass{JHEP3}

\usepackage{amsmath}
\usepackage{amsfonts}
\usepackage{amssymb}
\usepackage{citesort}
\usepackage{mathrsfs}

\def\S{{\mathbb S}}
\def\V{{\mathscr V}}

\newcommand{\alg}[1]{\mathfrak{#1}}
\newcommand{\su}{\alg{su}}

\def\ads{{\rm AdS}_5\times {\rm S}^5}

\author{Marius de Leeuw \footnote{Email: M.deLeeuw@uu.nl}
 \\  {\it Institute for Theoretical
Physics and Spinoza Institute,\\ Utrecht University, 3508 TD
Utrecht, The Netherlands}}

\abstract{We show that the recently found S-matrices  describing
the scattering of two-particle bound states of the light-cone
string sigma model on $\ads$ are compatible with Yangian symmetry.
In case the invariance with respect to the centrally extended
$\su(2|2)$ algebra is not sufficient to fully specify the
scattering matrix, the requirement of Yangian symmetry provides an
alternative to the Yang-Baxter equation and leads to a complete,
up to an overall phase, determination of the S-matrix. We then
compare the semi-classical limit of the bound state S-matrices
with the universal classical $r$-matrix by Beisert and Spill
evaluated in the corresponding bound state representations and
find perfect agreement. }

\title{Bound States, Yangian Symmetry and \\
Classical $r$-matrix for the $\ads$ Superstring }

\preprint{
          \tiny{ITP-UU-08-18}\\[-.5ex]
          \tiny{SPIN-08-17}\\[-.5ex]
          }
\begin{document}

\section{Introduction}

The discovery of integrable structures in the context of the
AdS/CFT correspondence \cite{Maldacena:1997re} has sparked many
new insights and developments in this field. It was first noted
that the operator spectrum of $\mathcal{N}=4$ super Yang-Mills
theory can be linked to (integrable) spin chains
\cite{Minahan:2002ve}. The classical string sigma model on $\ads$
was also shown to be integrable \cite{Bena:2003wd}. Although a
complete proof of integrability of the spin chain associated to
planar $\mathcal{N}= 4$ super Yang-Mills theory and of its string
dual is still missing, there is a lot of inspiring evidence that
integrability is indeed preserved. Assuming integrability to hold
has many important consequences. For example, the set of particle
momenta is conserved and every scattering process factorizes into
a sequence of two-body interactions. In other words, all the
scattering information is encrypted in the two-body
S-matrix.\smallskip

As in many physical theories, symmetry algebras also play a
crucial role here. The centrally extended $\su(2|2)$ superalgebra
has been shown to govern the asymptotic spectrum of the spin chain
associated to planar $\mathcal{N}=4$ super Yang-Mills theory at
higher loops \cite{Beisert:2004hm,Beisert:2005tm}. The very same
algebra also emerges from string theory \cite{Arutyunov:2006ak} as
a symmetry algebra of the light-cone Hamiltonian
\cite{Arutyunov:2004yx,Frolov:2006cc}. The requirement on the
S-matrix to be invariant under the centrally extended
$\mathfrak{su}(2|2)$ algebra fixes it uniquely up to an overall
phase factor \cite{Beisert:2005tm} and a choice of the
representation basis \cite{Arutyunov:2006yd}. With a proper choice
of the scattering basis, the S-matrix exhibits most of the
expected properties for a massive two-dimensional integrable field
theory, including unitarity and crossing symmetry. It also obeys
the Yang-Baxter equation \cite{Arutyunov:2006yd}.\smallskip

The S-matrix approach \cite{Serban:2004jf,Staudacher:2004tk} was
first developed in the spin chain framework of perturbative gauge
theory. It allowed one to conjecture the corresponding ``all-loop"
Bethe equations describing the gauge theory asymptotic spectrum
\cite{Beisert:2005fw,Beisert:2005tm}. On the string side, based on
the knowledge of the classical finite-gap solutions
\cite{Kazakov:2004qf}, a Bethe ansatz for the $\su(2)$ sector of
the  string sigma model was proposed \cite{Arutyunov:2004vx}. The
above mentioned non-analytic, overall (dressing) phase constitutes
an important feature of the string S-matrix. It has been a subject
of intensive research, see e.g.
\cite{Beisert:2006ib,Freyhult:2006vr,Eden:2006rx,Beisert:2007hz}.
Most importantly, by combining its expansion in terms of local
conserved charges with the requirement of crossing symmetry
\cite{Janik:2006dc}, one can find physically interesting solutions
\cite{Beisert:2006ib,Beisert:2006ez}, which nicely incorporate all
available string and gauge theory data. The algebraic and the
coordinate Bethe ans\"atze based on the string S-matrix have also
been studied in \cite{Martins:2007hb,Leeuw:2007uf}.\smallskip

The study of the quantum/classical scattering matrices and their
symmetries is also important for understanding finite size
effects. Away from the infinite volume(charge) limit, wrapping
interactions come into play and preclude the use of asymptotic
Bethe equations. So far there are two attempts to deal with this
problem. The first one consists in direct computation of
finite-size corrections to the giant magnon \cite{Hofman:2006xt}
(or bound state) dispersion relation  by using the
sigma-model/algebraic curve approach
\cite{Arutyunov:2006gs,Minahan:2008re,Klose:2008rx,Hatsuda:2008gd}.
Alternatively, these corrections can be obtained  by using
L\"uscher's perturbative approach
\cite{Janik:2007wt,Gromov:2008ie,Heller:2008at,Ambjorn:2005wa}.
The second way \cite{Arutyunov:2007tc} makes use  of the
thermodynamic Bethe ansatz \cite{Zamolodchikov:1989cf}. One can
define a mirror model \cite{Arutyunov:2007tc} for which the finite
size effects in the original theory are traded for finite
temperature effects in the infinite volume. Both the L\"uscher and
the TBA approaches rely on the knowledge of the corresponding
scattering matrices.
\smallskip

The two-body S-matrix that plays a pivotal role in the whole story
actually has an even larger symmetry algebra\footnote{See
\cite{Arutyunov:2006yd} for an earlier discussion of higher
symmetries of the fundamental S-matrix.} of Yangian type
 \cite{Beisert:2007ds,Matsumoto:2007rh}. Since Yangians have a number of
useful properties, in particular, at the level of representation
theory \cite{Bernard:1992ya,MacKay:2004tc}, appearance of Yangian
symmetry in the string context is quite a welcome feature. As a
matter of fact, the existence of Yangian symmetry gives hope of
constructing the universal R-matrix. Upon specifying suitable
representations, this R-matrix would then reproduce various
scattering processes; in particular, those involving the bound
states.\smallskip

At present, the existence of the universal R-matrix for the string
sigma model is an open problem. On the other hand, there are two
proposals for the classical $r$-matrix
\cite{Moriyama:2007jt,Beisert:2007ty}, which might arise in the
semi-classical limit of the yet to be found universal quantum
R-matrix. The second proposal \cite{Beisert:2007ty} was shown to
arise from the canonical $r$-matrix of the exceptional algebra
$\mathfrak{d}(2,1;\epsilon)$ which is closely related to the
$\su(2|2)$ algebra \cite{Matsumoto:2008ww}. From the string theory
point of view, the classical $r$-matrix corresponds to the
two-body S-matrix in the near plane-wave limit.
\smallskip

In addition to fundamental particles, the string sigma-model also
contains bound states \cite{Dorey:2006dq}. They fall into short
(atypical) symmetric representations of the centrally extended
$\mathfrak{su}(2|2)$ algebra
\cite{Dorey:2006dq,Beisert:2006qh,Chen:2006gp}. In a recent work
\cite{Arutyunov:2008zt} the S-matrices $\S^{AB}$ and $\S^{BB}$
which describe the scattering processes involving the fundamental
multiplet (A) and the two-particle bound state multiplet (B) have
been found. It appears that the extended $\mathfrak{su}(2|2)$
symmetry together with the Yang-Baxter equations is sufficient to
completely determine these S-matrices, up to an overall phase; the
overall phase can be chosen to satisfy the additional requirement
of crossing symmetry.
\smallskip

The aim of the present paper is to study the Yangian symmetry of
the bound state S-matrices from \cite{Arutyunov:2008zt} and also
to compare them to the proposed classical $r$-matrix from
\cite{Beisert:2007ty} in the near plane-wave limit. We find that
the S-matrices indeed respect Yangian symmetry. Moreover, as an
alternative to the Yang-Baxter equation, Yangian symmetry
completely determines the S-matrix $\S^{BB}$ up to a phase.
Finally, upon comparing the proposed classical $r$-matrix to the
bound state S-matrices in the near plane-wave limit, we find
perfect agreement.
\smallskip

The paper is organized as follows. First, we recall the structure
of the centrally extended $\mathfrak{su}(2|2)$ and the structure
of its Yangian. Subsequently, we will discuss the formulation of
the bound state representation in terms of differential operators
that we used for our computations. In this language we specify
coproducts of the $\su(2|2)$ symmetry generators and of the
Yangian generators and show that the bound state S-matrices
respect Yangian symmetry. Last, we discuss the classical $r$
matrix and compare it to the bound state S-matrices in the near
plane-wave limit.

\section{Centrally extended $\mathfrak{su}(2|2)$ and Yangians}

The algebra which plays a key role in the entire discussion is
centrally extended $\mathfrak{su}(2|2)$, which we will denote by
$\mathfrak{h}$. It is the symmetry algebra of the light-cone
Hamiltonian of the $\ads$ superstring and it also appears as the
symmetry algebra of the spin chain connected to $\mathcal{N}=4$
SYM. The algebra consists of bosonic generators
$\mathbb{R},\mathbb{L}$, supersymmetry generators
$\mathbb{Q},\mathbb{G}$ and central elements
$\mathbb{H},\mathbb{C},\mathbb{C}^{\dag}$. The non-trivial
commutation relations between the generators are given by
\begin{eqnarray}
\begin{array}{lll}
\ [\mathbb{L}_{a}^{\ b},\mathbb{J}_{c}] = \delta_{c}^{b}\mathbb{J}_{a}-\frac{1}{2}\delta_{a}^{b}\mathbb{J}_{c} &\qquad & \ [\mathbb{R}_{\alpha}^{\ \beta},\mathbb{J}_{\gamma}] = \delta_{\gamma}^{\beta}\mathbb{J}_{\alpha}-\frac{1}{2}\delta_{\alpha}^{\beta}\mathbb{J}_{\gamma}\\
\ [\mathbb{L}_{a}^{\ b},\mathbb{J}^{c}] = -\delta_{a}^{c}\mathbb{J}^{b}+\frac{1}{2}\delta_{a}^{b}\mathbb{J}^{c} &\qquad& \ [\mathbb{R}_{\alpha}^{\ \beta},\mathbb{J}^{\gamma}] = -\delta^{\gamma}_{\alpha}\mathbb{J}^{\beta}+\frac{1}{2}\delta_{\alpha}^{\beta}\mathbb{J}^{\gamma}\\
\ \{\mathbb{Q}_{\alpha}^{\ a},\mathbb{Q}_{\beta}^{\
b}\}=\epsilon_{\alpha\beta}\epsilon^{ab}\mathbb{C}&\qquad&\ \{\mathbb{G}^{\ \alpha}_{a},\mathbb{G}^{\ \beta}_{b}\}=\epsilon^{\alpha\beta}\epsilon_{ab}\mathbb{C}^{\dag}\\
\ \{\mathbb{Q}_{\alpha}^{a},\mathbb{G}^{\beta}_{b}\} =
\delta_{b}^{a}\mathbb{R}_{\alpha}^{\ \beta} +
\delta_{\alpha}^{\beta}\mathbb{L}_{b}^{\ a}
+\frac{1}{2}\delta_{b}^{a}\delta_{\alpha}^{\beta}\mathbb{H}.&&
\end{array}
\end{eqnarray}
The first two lines show how the indices of an arbitrary generator
with relevant indices transform. We will denote the eigenvalues of
the central charges by $H,C,C^{\dag}$. The charge $H$ is Hermitian
and the charges $C,C^{\dagger}$ are conjugate as well as the
generators $\mathbb{Q},\mathbb{G}$, i.e. $\mathbb{G}=
\mathbb{Q}^{\dagger}$.
\smallskip

Let us now turn our attention to the (double) Yangian of centrally
extended $\mathfrak{su}(2|2)$. We will briefly give the most
relevant definitions and results, and refer to
\cite{Bernard:1992ya,MacKay:2004tc} for more detailed accounts on
Yangians in general and to \cite{Beisert:2007ds,Beisert:2007ty}
for more details on the Yangian structure of $\mathfrak{h}$.

\subsubsection*{Double Yangian, generalities}
The double Yangian $DY(\mathfrak{g})$ of a (simple) Lie algebra
$\mathfrak{g}$ is a deformation of the universal enveloping
algebra $U(\mathfrak{g}[u,u^{-1}])$ of the loop algebra
$\mathfrak{g}[u,u^{-1}]$. Let us denote the deformation parameter
by $\hbar$. The Yangian is generated by level $n$ generators
$\mathbb{J}^{A}_{n},\ n\in\mathbb{Z}$ that satisfy the commutation
relations
\begin{eqnarray}
\ [\mathbb{J}^{A}_{m},\mathbb{J}^{B}_{n}  ]  = F^{AB}_{C} \mathbb{J}^{C}_{m+n} + \mathcal{O}(\hbar),
\end{eqnarray}
where $F^{AB}_{C}$ are the structure constants of $\mathfrak{g}$.
The level-0 generators $\mathbb{J}_{0}^{A}$ span the
Lie-algebra. The coproduct is given by
\begin{eqnarray}
\Delta (\mathbb{J}^{A}_{n})  = \mathbb{J}^{A}_{n}\otimes 1 +  1\otimes \mathbb{J}^{A}_{n} + \frac{\hbar }{2} \sum_{m=0}^{n-1} F_{BC}^{A}
\mathbb{J}^{B}_{n-1-m}\otimes \mathbb{J}^{C}_{m}.
\end{eqnarray}
Where the indices on the structure constants were lowered with the
Cartan-Killing matrix.\smallskip

The Yangian can be supplied with the structure of a
quasi-cocommutative Hopf-algebra if there is an R-matrix, $R\in
DY(\mathfrak{g})\otimes DY(\mathfrak{g})$ such that
\begin{eqnarray}\label{eqn;CoComm}
\Delta^{op}(\mathbb{J}^{A}_{n}) R  = R \Delta (\mathbb{J}^{A}_{n}),
\end{eqnarray}
with $\Delta^{op}$ the opposite coproduct, $\Delta^{op}= P\Delta
$, where $P$ is the (graded) permutation operator. For
conventional Yangians this universal R-matrix exists and can be
explicitly constructed with the help of the Cartan-Killing
matrix.\smallskip

An important representation of the Yangian is the evaluation
representation. This representation consists of states
$|u\rangle$, with action $\mathbb{J}^{A}_{n}|u\rangle =
u^{n}\mathbb{J}^{A}_{0}|u\rangle$. Hence, upon choosing a
representation of the Lie algebra we obtain a representation of
the Yangian. The coproduct becomes particulary easy in this
representation. Let it act on the state
$|u_{1}\rangle\otimes|u_{2}\rangle$, then it is of the form:
\begin{eqnarray}\label{eqn;CoProdRmat}
\Delta(\mathbb{J}^{A}_{n})  \approx \frac{u_{1}^{n-1}-u_{2}^{n-1}}{u_{1}^{-1}-u_{2}^{-1}} \Delta(\mathbb{J}_{0}^{A})+ \frac{u_{1}^{n}-u_{2}^{n}}{u_{1}-u_{2}} \Delta(\mathbb{J}_{1}^{A}).
\end{eqnarray}
This means that if one wants to check invariance of an R-matrix
under Yangian symmetry in the evaluation representation it is
enough to check this for
$\mathbb{J}_{0}^{A},\mathbb{J}_{1}^{A}$.\smallskip

The parameter $\hbar$ is viewed as a quantum parameter and the
Yangian as a quantum deformation of the enveloping algebra. We can
consider the semi-classical limit by working consistently up to
order $\hbar$. In this limit, the universal R-matrix expands as:
\begin{eqnarray}\label{eqn;ClasRmat}
R = 1 + \hbar r + \mathcal{O}(\hbar^{2}).
\end{eqnarray}
The operator $r$ is called the classical $r$-matrix.

\subsubsection*{Double Yangian, centrally extended $\mathfrak{su}(2|2)$}

Unfortunately $\mathfrak{h}$ is not simple and hence the above
discussed methods cannot be straightforwardly applied.

For the coproduct one needs to introduce a non-trivial braiding
\cite{Gomez:2006va,Plefka:2006ze,Beisert:2007ds},
\begin{eqnarray}\label{eqn;CoProdYang}
\Delta (\mathbb{J}^{A}_{n})  &=& \mathbb{J}^{A}_{n}\otimes 1 +  \mathcal{U}^{[A]}\otimes \mathbb{J}^{A}_{n} + \frac{\hbar }{2} \sum_{m=0}^{n-1} F_{BC}^{A}
\mathbb{J}^{B}_{n-1-m}\mathcal{U}^{[C]}\otimes \mathbb{J}^{C}_{m} +\mathcal{O}(\hbar^{2})\nonumber\\
\Delta(\mathcal{U})&=&\mathcal{U}\otimes \mathcal{U},
\end{eqnarray}
for some Abelian generator $\mathcal{U}$ and ``braid charges'':
\begin{eqnarray}
[\mathbb{C}^{\dag}]=-2,\quad[\mathbb{G}]=-1,\quad[\mathbb{L}]=[\mathbb{R}]=[\mathbb{H}]=0,\quad[\mathbb{Q}]=1,\quad
[\mathbb{C}]=2.
\end{eqnarray}
In section \ref{sec;TPcPS-mat} we will use another formulation of
the coproduct which avoids the explicit use of the braiding
factors. In order to make supply the Yangian with a
quasi-cocommutative structure, it was shown that the central
charges $\mathbb{C},\mathbb{C}^{\dag}$ need to be identified with
the braiding factor $\mathcal{U}$ and $\mathbb{H}$ in the
following way \cite{Plefka:2006ze,Beisert:2007ds}:
\begin{eqnarray}
\begin{array}{lll}
\mathbb{C}_{0}\sim g(1-\mathcal{U}^2) &\quad &\mathbb{C}^{\dag}_{0} \sim g(1-\mathcal{U}^{-2})\\
\mathbb{C}_{1}\sim \mathbb{H}(1+\mathcal{U}^2)&\quad
&\mathbb{C}^{\dag}_{1} \sim -\mathbb{H}(1+\mathcal{U}^{-2}).
\end{array}
\end{eqnarray}
In the evaluation representation we can use this to express the
evaluation parameter $u$ in terms of the eigenvalues of $H$ and
the braiding operator:
\begin{eqnarray}
\mathbb{J}^{A}_{n} = (iu)^{n}\mathbb{J}^{A}_{0},\qquad iu \sim
 \mathbb{H}\frac{1+\mathcal{U}^{2}}{1-\mathcal{U}^{2}}
\end{eqnarray}
This is quite different from the standard case, where the
evaluation parameter is unrelated to the algebra. The adjusted
notion of coproduct raises the question whether a universal
R-matrix can still be found in order to make the Yangian
quasi-cocommutative. One could hope that, just as in the simple
case, it could be constructed by means of the Cartan-Killing
matrix. However, for $\mathfrak{h}$ this appears to be singular
and hence the standard construction breaks down. Nevertheless, for
the fundamental evaluation represenation of $\mathfrak{h}$, the
R-matrix has been found as a scattering matrix
\cite{Beisert:2005tm,Arutyunov:2006yd} and it indeed respects
Yangian symmetry \cite{Beisert:2007ds}.\smallskip

Although an expression for the universal R-matrix is currently
lacking, there have been proposals for the classical $r$-matrix
\cite{Moriyama:2007jt,Beisert:2007ty}. We will focus on the
proposal \cite{Beisert:2007ty} in section 4.

\section{Representation with differential operators and symmetry invariance}

The representations that describe $M$-particle bound states are
$4M$-dimensional and because of this, the sizes of the involved
matrices quickly get out of hand. To avoid doing computations with
unwieldy matrices it is useful to put this representation in the
formalism of differential operators. The encountered totally
symmetric representation of $M$-particle bound states can be
identified with a $4M$-dimensional graded vector space of
monomials of degree $M$ and the different generators can be
represented by corresponding differential operators
\cite{Arutyunov:2008zt}.

\subsection{Formalism and bound state representations}

Consider the vector space of analytic functions of two bosonic
variables $w_{a}$ and two fermionic variables $\theta_{\alpha}$.
Since we are dealing with analytic functions we can expand any
such function $\Phi(w,\theta)$:
\begin{eqnarray}
\Phi(w,\theta) &=&\sum_{M=0}^{\infty}\Phi_{M}(w,\theta),\nonumber\\
\Phi_{M} &=& \phi^{a_{1}\ldots a_{M}}w_{a_{1}}\ldots w_{a_{M}}
+\phi^{a_{1}\ldots a_{M-1}\alpha}w_{a_{1}}\ldots
w_{a_{M-1}}\theta_{\alpha}+\nonumber\\
&&\phi^{a_{1}\ldots a_{M-2}\alpha\beta}w_{a_{1}}\ldots
w_{a_{M-2}}\theta_{\alpha}\theta_{\beta}.
\end{eqnarray}
The representation of centrally extended $\mathfrak{su}(2|2)$,
that describes $M$-particle bound states of the light-cone string
theory has dimension $4M$. It is realized on a graded vector space
with basis $|e_{a_{1}\ldots a_{M}}\rangle, |e_{a_{1}\ldots
a_{M-1}\alpha}\rangle,|e_{a_{1}\ldots a_{M-2}\alpha\beta}\rangle$,
where $a_{i}$ are bosonic indices and $\alpha,\beta$ are fermionic
indices and each of the basis vectors is totally symmetric in the
bosonic indices and anti-symmetric in the fermionic indices
\cite{Arutyunov:2008zt,Beisert:2006qh,Dorey:2006dq}. In terms of
the above analytic functions, the basis vectors of the totally
symmetric representation can evidently be identified
$|e_{a_{1}\ldots a_{M}}\rangle \leftrightarrow w_{a_{1}}\ldots
w_{a_{M}},|e_{a_{1}\ldots a_{M-1}\alpha}\rangle \leftrightarrow
w_{a_{1}}\ldots w_{a_{M-1}}\theta_{\alpha}$ and $|e_{a_{1}\ldots
a_{M-1}\alpha\beta}\rangle \leftrightarrow w_{a_{1}}\ldots
w_{a_{M-2}}\theta_{\alpha}\theta_{\beta}$ respectively.  In other
words, we find the atypical totally symmetric representation
describing $M$-particle bound states when we restrict to terms $
\Phi_{M}$.
\smallskip

In this representation the generators of $\mathfrak{h}$ can be
written in differential operator form in the following way
\begin{eqnarray}\label{eqn;AlgDiff}
\begin{array}{lll}
  \mathbb{L}_{a}^{\ b} = w_{a}\frac{\partial}{\partial w_{b}}-\frac{1}{2}\delta_{a}^{b}w_{c}\frac{\partial}{\partial w_{c}}, &\qquad& \mathbb{R}_{\alpha}^{\ \beta} = \theta_{\alpha}\frac{\partial}{\partial \theta_{\beta}}-\frac{1}{2}\delta_{\alpha}^{\beta}\theta_{\gamma}\frac{\partial}{\partial \theta_{\gamma}}, \\
  \mathbb{Q}_{\alpha}^{\ a} = a \theta_{\alpha}\frac{\partial}{\partial w_{a}}+b\epsilon^{ab}\epsilon_{\alpha\beta} w_{b}\frac{\partial}{\partial \theta_{\beta}}, &\qquad& \mathbb{G}_{a}^{\ \alpha} = d w_{a}\frac{\partial}{\partial \theta_{\alpha}}+c\epsilon_{ab}\epsilon^{\alpha\beta} \theta_{\beta}\frac{\partial}{\partial w_{b}}
\end{array}
\end{eqnarray}
and the central charges are
\begin{eqnarray}
\begin{array}{ll}
 \mathbb{C} = ab \left(w_{a}\frac{\partial}{\partial w_{a}}+\theta_{\alpha}\frac{\partial}{\partial
 \theta_{\alpha}}\right)& \mathbb{C}^{\dag} = cd \left(w_{a}\frac{\partial}{\partial w_{a}}+\theta_{\alpha}\frac{\partial}{\partial
 \theta_{\alpha}}\right)\\
 \mathbb{H}= (ad +bc)\left(w_{a}\frac{\partial}{\partial w_{a}}+\theta_{\alpha}\frac{\partial}{\partial
 \theta_{\alpha}}\right).
\end{array}
\end{eqnarray}
To form a representations, the parameters $a,b,c,d$ satisfy the
condition $ad-bc=1$. The central charges become $M$ dependent:
\begin{eqnarray}
H= M(ad+bc),\qquad C =M ab , \qquad C^{\dag} =M cd .
\end{eqnarray}

In what follows we will also need an additional operator
\footnote{We are grateful to G. Arutyunov for suggesting this.}
\begin{eqnarray}
\Sigma =\frac{1}{2} \frac{1}{ad+bc}\left(
w_{a}\frac{\partial}{\partial w_{a}}
-\theta_{a}\frac{\partial}{\partial\theta_{\alpha}}\right).
\end{eqnarray}
This operator corresponds (up to the prefactor) to the grading
matrix $\Sigma$ of \cite{Arutyunov:2006yd} and it distinguishes
the superfield components with different numbers of fermions. On
bound state representations $\Sigma$ has the following commutation
relations with the algebra generators
\begin{eqnarray}
\ [\Sigma,\mathbb{Q}^{\alpha}_{\ b} ] &=&- \mathbb{Q}^{\alpha}_{\
b} + 2
\mathbb{C}\mathbb{H}^{-1}\epsilon^{\alpha\gamma}\epsilon_{bd}
\mathbb{G}^{d}_{\
\gamma}\nonumber\\
\ [\Sigma,\mathbb{G}^{a}_{\ \beta} ] &=& \mathbb{G}^{a}_{\ \beta}
- 2\mathbb{C}^{\dagger}\mathbb{H}^{-1}
\epsilon_{\beta\gamma}\epsilon^{ad}
\mathbb{Q}^{\gamma}_{\ d}\\
\ [\Sigma,\mathbb{L}^{a}_{\ b} ]&=&\
[\Sigma,\mathbb{R}^{\alpha}_{\ \beta} ]=\ [\Sigma,\mathbb{H}
]=0\nonumber.
\end{eqnarray}
We will also introduce the following quadratic operator
\begin{eqnarray}
\mathcal{T}= \mathbb{R}^{\ \alpha}_{\beta}\mathbb{R}^{\
\beta}_{\alpha}- \mathbb{L}^{\ a}_{b}\mathbb{L}^{\ b}_{a}+
\mathbb{G}^{\ a}_{\alpha}\mathbb{Q}^{\ \alpha}_{a}- \mathbb{Q}^{\
\alpha}_{a}\mathbb{G}^{\ a}_{\alpha}.
\end{eqnarray}
The operators $\Sigma$ and $\mathcal{T}$ can be used to construct
the Casimir operator $\mathfrak{C}$ of the $\mathfrak{u}(2|2)$
algebra
\begin{eqnarray}\label{eqn;Casimir}
\mathfrak{C} = \Sigma\mathbb{H} - \mathcal{T}
\end{eqnarray}
which in the $M$-particle bound state representation has the
following eigenvalue
\begin{eqnarray}
\mathfrak{C}|M\rangle = M(M-1)|M\rangle\, .
\end{eqnarray}

Further, we introduce the parameterization for $a,b,c,d$ in terms
of the particle momentum and the coupling $g$:
\begin{eqnarray}
\begin{array}{lll}
  a = \sqrt{\frac{g}{2M}}\eta & \quad & b = \sqrt{\frac{g}{2M}}
\frac{i\zeta}{\eta}\left(\frac{x^{+}}{x^{-}}-1\right) \\
  c = -\sqrt{\frac{g}{2M}}\frac{\eta}{\zeta x^{+}} & \quad &
d=\sqrt{\frac{g}{2M}}\frac{x^{+}}{i\eta}\left(1-\frac{x^{-}}{x^{+}}\right),
\end{array}
\end{eqnarray}
where the parameters $x^{\pm}$ satisfy
\begin{eqnarray}
x^{+} +
\frac{1}{x^{+}}-x^{-}-\frac{1}{x^{-}}=\frac{2Mi}{g},\qquad
\frac{x^{+}}{x^{-}} = e^{ip}.
\end{eqnarray}
Finally, the eigenvalue of the braid operator is found to be
$\mathcal{U} = \sqrt{\frac{x^{+}}{x^{-}}}$ and the parameter of
the evaluation representation is identified with $u_{j}=
x_{j}^{+}+\frac{1}{x_{j}^{+}} - \frac{iM_{j}}{g}$. The fundamental
representation corresponds to taking $M=1$.\smallskip

The totally symmetric representation is now completely fixed by
specifying $x^{\pm},g,\eta,M$. The factor of $\eta$ reflects a
freedom in choosing basis vectors. However, as found in
\cite{Arutyunov:2007tc}, it appears that string theory selects a
particular choice of $\eta,\zeta$:
\begin{eqnarray}
\eta = e^{i\xi}e^{\frac{i}{4}p}\sqrt{ix^{-}-ix^{+}},\qquad
\zeta = e^{2i\xi}.
\end{eqnarray}
As a consequence of this choice, the S-matrix satisfies the
normal, untwisted Yang-Baxter equation and is, in fact, a
symmetric operator. Adopting this choice also has some
consequences for the braiding factor in the coproduct. This will
be discussed in the next sections.

\subsection{Tensor products, coproducts and symmetries}\label{sec;TPcPS-mat}

When analyzing R-matrices or S-matrices one needs to consider
(graded) tensor products of representations. In the context of
differential operators in a superspace, this is easily realized by
considering the product of two irreducible superfields
$\Phi_{M_{1}}(w_{a},\theta_{\alpha})\Phi_{M_{2}}(u_{a},\vartheta_{\alpha})$
depending on different sets of coordinates.\smallskip

\subsubsection*{Coproduct of the algebra generators}

Let us now consider the coproducts of the symmetry generators of
centrally extended $\su(2|2)$. As discussed in
\cite{Arutyunov:2006yd}, the S-matrix is a map between the
following representations:
\begin{eqnarray}
\S:~~ \V_{M_{1}}(p_{1},e^{ip_{2}})\otimes \V_{M_{2}}(p_{2},1
)\longrightarrow \V_{M_{1}}(p_{1},1 )\otimes
\V_{M_{2}}(p_{2},e^{ip_{1}}),
\end{eqnarray}
where $\V_{M_{i}}(p_{i},e^{2i\xi})$ is the $M_{i}$-bound state
representation with parameters $a_{i},b_{i},c_{i},d_{i}$ with the
explicit choice of $\zeta = e^{2i\xi}$. Taking into account the
above parameters of the different representations, we see that
when checking the relation
\begin{eqnarray}
\S~\Delta(\mathbb{J}_{0}^{A})=\Delta^{op}(\mathbb{J}_{0}^{A})~\S,
\end{eqnarray}
all the explicit braiding factors drop out and we get
\begin{eqnarray}\label{eqn;CoprodSymm}
\Delta(\mathbb{J}_{0}^{A}) =\mathbb{J}_{1;0}^{A} +
\mathbb{J}_{2;0}^{A}.
\end{eqnarray}
Here $\Delta^{op}(\mathbb{J}_{i}^{A})$ acts on $\V_{M_{1}}(p_{1},1
)\otimes \V_{M_{2}}(p_{2},e^{ip_{1}})$ and
$\Delta(\mathbb{J}_{i}^{A})$ acts on
$\V_{M_{1}}(p_{1},e^{ip_{2}})\otimes \V_{M_{2}}(p_{2},1 )$, with
the appropriate coefficients $a,b,c,d$. We will give explicit
expressions in the next section for each of these coefficients. In
the above formula $\mathbb{J}_{i;0}^{A}$ is the operator
$\mathbb{J}_{0}^{A}$ acting in the $i$-th space.\smallskip

In \cite{Beisert:2005tm,Arutyunov:2006yd} the requirement of
invariance under $\mathfrak{h}$, in the sense of commuting with
(\ref{eqn;CoprodSymm}), was found to be sufficient to fix the
fundamental S-matrix, denoted by $\S^{AA}$, up to a phase factor.
This procedure has also been carried out for the two-particle
bound states representations \cite{Arutyunov:2008zt}. This,
together with Yang-Baxter, was again enough to fix the involved
S-matrices up to a phase factor. For the explicit form of these
S-matrices, denoted by $\S^{AB}$ and $\S^{BB}$, we refer to
\cite{Arutyunov:2008zt}.\smallskip

We will spell out the phase factors, since they will play a role
when comparing the S-matrices to the classical $r$-matrix. For
$\S^{AA}$ the corresponding phase factor has been studied quite
intensively. This factor allows one to derive the phase factors
for $\S^{AB}$ and $\S^{BB}$ by applying the fusion
procedure\footnote{The fusion procedure for rational S/R-matrices
based on $\mathfrak{gl}(m|n)$ has recently been worked out
\cite{Kazakov:2007na}.}
\cite{Dorey:2006dq,Roiban:2006gs,Chen:2006gq}. Define the function
\begin{eqnarray}
G(n):=\frac{u_{1}-u_{2}+\frac{in}{g}}{u_{1}-u_{2}-\frac{in}{g}},
\end{eqnarray}
where $u_{j} = x_{j}^{+}+\frac{1}{x_{j}^{+}} - \frac{iM_{j}}{g}$. The
phase factors of the different matrices that follow from fusion
and crossing symmetry are:
\begin{eqnarray}\label{eqn;Phase}
S^{AA}_{0}&=&\sqrt{G(0)G(2)}~\sqrt{\frac{x_{1;1}^{-}x_{2;1}^{+}}{x_{1;1}^{+}x_{2;1}^{-}}}~\sigma(x_{1;1},x_{1;1})\nonumber\\
S^{AB}_{0}&=&\sqrt{G(1)G(3)}~\frac{x_{1;2}^{-}}{x_{1;2}^{+}}\sqrt{\frac{x_{2;2}^{+}}{x_{2;2}^{-}}}~\sigma(x_{1;1},x_{2;2})\\
S^{BB}_{0}&=&G(2)\sqrt{G(4)}~\frac{x_{1;2}^{-}x_{2;2}^{+}}{x_{1;2}^{+}x_{2;2}^{-}}~\sigma(x_{1;2},x_{2;2}),\nonumber
\end{eqnarray}
where $\sigma(p,q) = e^{i\theta(p,q)}$ is the dressing phase. The
canonical S-matrices, completed with the above phases respect
crossing symmetry \cite{Arutyunov:2008zt}.

\subsubsection*{Yangian symmetry of the S-matrices}
It appears that the coproduct of the Yangian generators can be
written in the standard form. Let us denote the first Yangian
generator of an operator $\mathbb{J}$ by $\hat{\mathbb{J}}$. All
the braiding factors entering in the coproduct
(\ref{eqn;CoProdYang}) never explicitly appear in the operator
formalism discussed above. The coproduct in the differential
operator form is now given by
\begin{eqnarray}\label{eqn;CoProdYangDiff}
\Delta(\hat{\mathbb{L}}^{a}_{\ b}) &=& \hat{\mathbb{L}}^{\
a}_{1;b} + \hat{\mathbb{L}}^{\ a}_{2;b} +
\frac{1}{2}\mathbb{L}^{\ a}_{1;c}\mathbb{L}^{\
c}_{2;b}-\frac{1}{2} \mathbb{L}^{\ c}_{1;b}\mathbb{L}^{\
a}_{2;c}-\frac{1}{2} \mathbb{G}^{\ a}_{1;\gamma}\mathbb{Q}^{\
\gamma}_{2;b}-\frac{1}{2} \mathbb{Q}^{\
\gamma}_{1;b}\mathbb{G}^{\ a}_{2;\gamma}\nonumber\\
&&+\frac{1}{4}\delta^{a}_{b}\mathbb{G}^{\
c}_{1;\gamma}\mathbb{Q}^{\
\gamma}_{2;c}+\frac{1}{4}\delta^{a}_{b} \mathbb{Q}^{\
\gamma}_{1;c}\mathbb{G}^{\ c}_{2;\gamma}\nonumber\\
\Delta(\hat{\mathbb{R}}^{\alpha}_{\ \beta}) &=&
\hat{\mathbb{R}}^{\ \alpha}_{1;\beta} + \hat{\mathbb{R}}^{\
\alpha}_{2;\beta} - \frac{1}{2}\mathbb{R}^{\
\alpha}_{1;\gamma}\mathbb{R}^{\ \gamma}_{2;\beta}+\frac{1}{2}
\mathbb{R}^{\ \gamma}_{1;\beta}\mathbb{R}^{\
\alpha}_{2;\gamma}+\frac{1}{2} \mathbb{G}^{\
c}_{1;\beta}\mathbb{Q}^{\ \alpha}_{2;c}+\frac{1}{2}
\mathbb{Q}^{\
\alpha}_{1;c}\mathbb{G}^{\ c}_{2;\beta}\nonumber\\
&&-\frac{1}{4}\delta^{\alpha}_{\beta}\mathbb{G}^{\
c}_{1;\gamma}\mathbb{Q}^{\
\gamma}_{2;c}-\frac{1}{4}\delta^{\alpha}_{\beta} \mathbb{Q}^{\
\gamma}_{1;c}\mathbb{G}^{\ c}_{2;\gamma}\\
\Delta(\hat{\mathbb{Q}}^{\alpha}_{\ b}) &=&
\hat{\mathbb{Q}}^{\ \alpha}_{1;b} + \hat{\mathbb{Q}}^{\
\alpha}_{2;b} - \frac{1}{2}\mathbb{R}^{\
\alpha}_{1;\gamma}\mathbb{Q}^{\ \gamma}_{2;b}+\frac{1}{2}
\mathbb{Q}^{\ \gamma}_{1;b}\mathbb{R}^{\ \alpha}_{2;\gamma}
-\frac{1}{2} \mathbb{L}^{\ c}_{1;b}\mathbb{Q}^{\
\alpha}_{2;c}+\frac{1}{2} \mathbb{Q}^{\
\alpha}_{1;c}\mathbb{L}^{\ c}_{2;b}\nonumber\\
&&-\frac{1}{4}\mathbb{H}_{1}\mathbb{Q}^{\
\alpha}_{2;b}+\frac{1}{4}\mathbb{Q}^{\ \alpha}_{1;b}\mathbb{H}_{2}
+
\frac{1}{2}\epsilon^{\alpha\gamma}\epsilon_{bd}\mathbb{C}_{1}\mathbb{G}^{\
d}_{2;\gamma}-\frac{1}{2}\epsilon^{\alpha\gamma}\epsilon_{bd}\mathbb{G}^{\
d}_{1;\gamma}\mathbb{C}_{2}\nonumber\\
\Delta(\hat{\mathbb{G}}^{a}_{\ \beta}) &=& \hat{\mathbb{G}}^{\
a}_{1;\beta} + \hat{\mathbb{G}}^{\ a}_{2;\beta} +
\frac{1}{2}\mathbb{L}^{\ a}_{1;c}\mathbb{G}^{\ c}_{2;\beta}-
\frac{1}{2}\mathbb{G}^{\ c}_{1;\beta}\mathbb{L}^{\ a}_{2;c}
+\frac{1}{2} \mathbb{R}^{\ \gamma}_{1;\beta}\mathbb{G}^{\
a}_{2;\gamma} +\frac{1}{2} \mathbb{G}^{\
a}_{1;\gamma}\mathbb{R}^{\ \gamma}_{2;\beta}\nonumber\\
&&+\frac{1}{4}\mathbb{H}_{1}\mathbb{G}^{\ a}_{2;\beta}+
\frac{1}{4}\mathbb{G}^{\ a}_{1;\beta}\mathbb{H}_{2} -
\frac{1}{2}\epsilon^{ac}\epsilon_{\beta\gamma}\mathbb{C}^{\dag}_{1}\mathbb{Q}^{\
\gamma}_{2;c}
+\frac{1}{2}\epsilon^{ac}\epsilon_{\beta\gamma}\mathbb{Q}^{\
\gamma}_{1;c}\mathbb{C}^{\dag}_{2}\nonumber
\end{eqnarray}
and for the central charges:
\begin{eqnarray}
\Delta(\hat{\mathbb{H}}) &=& \hat{\mathbb{H}}_{1} +\hat{\mathbb{H}}_{2}+\frac{1}{2} \mathbb{C}_{1}\mathbb{C}^{\dag}_{2}-\frac{1}{2} \mathbb{C}^{\dag}_{1}\mathbb{C}_{2}\nonumber\\
\Delta(\hat{\mathbb{C}}) &=& \hat{\mathbb{C}}_{1}+\hat{\mathbb{C}}_{2}+\frac{1}{2} \mathbb{H}_{1}\mathbb{C}_{2}-\frac{1}{2} \mathbb{C}_{1}\mathbb{H}_{2}\\
\Delta(\hat{\mathbb{C}}^{\dag}) &=&
\hat{\mathbb{C}}^{\dag}_{1}+\hat{\mathbb{C}}^{\dag}_{2}+\frac{1}{2}
\mathbb{H}_{1}\mathbb{C}^{\dag}_{2}-\frac{1}{2}
\mathbb{C}^{\dag}_{1}\mathbb{H}_{2}\nonumber.
\end{eqnarray}
The product is ordered, e.g. $\mathbb{Q}_{1}\mathbb{Q}_{2}$ means
first apply $\mathbb{Q}_{2}$, then $\mathbb{Q}_{1}$. Also, in the
evaluation representation we identify $\hat{\mathbb{J}} =
\frac{g}{2i}u\mathbb{J}$. As stressed in the previous section,
$\Delta^{op}$ acts on representations with different parameters
$\zeta$. For completeness we will explicitly give the parameters
$a,b,c,d$ for the involved representations. The coefficients for
$\Delta(\mathbb{J})$ are given by:
\begin{eqnarray}
\begin{array}{lll}
  a_{1} = \sqrt{\frac{g}{2M_{1}}}\eta_{1} & ~ & b_{1} =
 -i e^{i p_{2}}\sqrt{\frac{g}{2M_{1}}}~
\frac{1}{\eta_{1}}\left(\frac{x_{1}^{+}}{x_{1}^{-}}-1\right) \\
  c_{1} = -e^{-i p_{2}}\sqrt{\frac{g}{2M_{1}}}\frac{\eta_{1}}{ x_{1}^{+}} & ~ &
d_{1}=i\sqrt{\frac{g}{2M_{1}}}\frac{x_{1}^{+}}{\eta_{1}}\left(\frac{x_{1}^{-}}{x_{1}^{+}}-1\right)\\
\eta_{1} =
e^{i\frac{p_{1}}{4}}e^{i\frac{p_{2}}{2}}\sqrt{ix^{-}_{1}-ix^{+}_{1}}&~&~\\
~ &~& ~ \\
  a_{2} = \sqrt{\frac{g}{2M_{2}}}\eta_{2} & ~ & b_{2} = -i\sqrt{\frac{g}{2M_{2}}}
\frac{1}{\eta_{2}}\left(\frac{x_{2}^{+}}{x_{2}^{-}}-1\right) \\
  c_{2} = -\sqrt{\frac{g}{2M_{2}}}\frac{\eta_{2}}{x_{2}^{+}} & ~ &
d_{2}=i\sqrt{\frac{g}{2M_{2}}}\frac{x_{2}^{+}}{i\eta_{2}}\left(\frac{x_{2}^{-}}{x_{2}^{+}}-1\right)\\
\eta_{2} = e^{i\frac{p_{2}}{4}}\sqrt{ix^{-}_{2}-ix^{+}_{2}}&~&~
\end{array}
\end{eqnarray}
The coefficients in $\Delta^{op}(\mathbb{J})$ are given by:
\begin{eqnarray}
\begin{array}{lll}
  a^{op}_{1} = \sqrt{\frac{g}{2M_{1}}}\eta^{op}_{1} & ~ & b^{op}_{1} = -i\sqrt{\frac{g}{2M_{1}}}
\frac{1}{\eta^{op}_{1}}\left(\frac{x_{1}^{+}}{x_{1}^{-}}-1\right) \\
  c^{op}_{1} = -\sqrt{\frac{g}{2M_{1}}}\frac{\eta^{op}_{1}}{x_{1}^{+}} & ~ &
d^{op}_{1}=i\sqrt{\frac{g}{2M_{1}}}\frac{x_{1}^{+}}{i\eta^{op}_{1}}\left(\frac{x_{1}^{-}}{x_{1}^{+}}-1\right)\\
\eta^{op}_{1} =
e^{i\frac{p_{1}}{4}}\sqrt{ix^{-}_{1}-ix^{+}_{1}}&~&~\nonumber\\
 ~ &~& ~ \\
  a^{op}_{2} = \sqrt{\frac{g}{2M_{2}}}\eta^{op}_{2} & ~ &
b^{op}_{2} =
 -i e^{i p_{1}}\sqrt{\frac{g}{2M_{2}}}~
\frac{1}{\eta^{op}_{2}}\left(\frac{x_{2}^{+}}{x_{2}^{-}}-1\right) \\
  c^{op}_{2} = -e^{-i p_{1}}\sqrt{\frac{g}{2M_{2}}}\frac{\eta^{op}_{2}}{ x_{2}^{+}} & ~ &
d^{op}_{2}=i\sqrt{\frac{g}{2M_{2}}}\frac{x_{2}^{+}}{\eta^{op}_{2}}\left(\frac{x_{2}^{-}}{x_{2}^{+}}-1\right)\\
\eta^{op}_{2} =
e^{i\frac{p_{2}}{4}}e^{i\frac{p_{1}}{2}}\sqrt{ix^{-}_{2}-ix^{+}_{2}}&~&~
\end{array}
\end{eqnarray}
According to the logic of \cite{Arutyunov:2006yd}, the non-trivial
braiding factors present in eq.(\ref{eqn;CoProdYang}) are all
hidden in the parameters of the four representations
involved.\smallskip

Using the above described differential representation, we have
verified that both $\S^{AB}$ and $\S^{BB}$ are invariant under
Yangian symmetry by explicitly showing that it cocommutes with the
above specified coproduct for Yangian generators
$\hat{\mathbb{J}}^{A}$:
\begin{eqnarray}
\S~\Delta(\hat{\mathbb{J}}^{A})=\Delta^{op}(\hat{\mathbb{J}}^{A})~\S.
\end{eqnarray}
Since we work in the evalutation representation, by
(\ref{eqn;CoProdRmat}), this indeed suffices. We omit the details
of the computation since they are not very illuminating.
\smallskip

Most importantly, Yangian symmetry fixes $\S^{BB}$ uniquely up to
phase factor without usage of the Yang-Baxter equation. In
\cite{Arutyunov:2008zt} it was found that by requiring $\S^{BB}$
to be invariant under $\mathfrak{h}$ fixes it up to two
coefficients (one being the overall phase which we omit here):
\begin{eqnarray}\label{eqn;SBBpreYB}
\S^{BB} = \S^{BB}_f+q \, \S^{BB}_s.
\end{eqnarray}
The coefficient was then determined by demanding $\S^{BB}$ to
satisfy the Yang-Baxter equation  \cite{Arutyunov:2008zt}. Here,
by insisting that the S-matrix (\ref{eqn;SBBpreYB}) respects
Yangian symmetry we found that this fixes $q$ uniquely and that
its value coincides with the one obtained in
\cite{Arutyunov:2008zt}. This feature of the higher (Yangian)
symmetries of the S-matrices as being a substitute for the
Yang-Baxter equation is not unexpected, as was explained in
\cite{Arutyunov:2006yd}. Our computation confirms this point and
simultaneously provides an independent check of the results by
\cite{Arutyunov:2008zt}.

\section{The near plane-wave limit and the classical $r$-matrix}

We will now concentrate on the plane-wave limit of the bound state
S-matrices. In this limit it should agree with the universal classical
$r$-matrix.

\subsection{The universal classical $r$-matrix}

In \cite{Beisert:2007ty} a proposal for the classical $r$-matrix
was made in terms of algebra generators in the evaluation
representation which in the classical limit coincides with the
S-matrix found in \cite{Beisert:2005tm,Arutyunov:2006yd}.
\smallskip

The $r$-matrix is completely given in terms of algebra generators
and evaluation parameters $u_{1},u_{2}$. Consider the following
two-site operator
\begin{eqnarray}
\mathcal{T}_{12}=2\left(\mathbb{R}^{\
\alpha}_{\beta}\otimes\mathbb{R}^{\ \beta}_{\alpha}- \mathbb{L}^{\
a}_{b}\otimes\mathbb{L}^{\ b}_{a}+ \mathbb{G}^{\
\alpha}_{a}\otimes\mathbb{Q}^{\ a}_{\alpha}- \mathbb{Q}^{\
a}_{\alpha}\otimes\mathbb{G}^{\ \alpha}_{a}\right).
\end{eqnarray}
Next we introduce an operator $\mathbb{B}$, which is subject to
the following relations (in the classical limit)
\begin{eqnarray}
\ [\mathbb{B}_{m},(\mathbb{Q}_{n})^{\alpha}_{\ b} ] &=&-
(\mathbb{Q}_{m+n})^{\alpha}_{\ b} + 2
\epsilon^{\alpha\gamma}\epsilon_{bd} (\mathbb{G}_{m+n-1})^{d}_{\
\gamma}\nonumber\\
\label{Bcom} \ [\mathbb{B}_{m},(\mathbb{G}_{n})^{a}_{\ \beta} ]
&=& (\mathbb{G}_{m+n})^{a}_{\ \beta} - 2
\epsilon_{\beta\gamma}\epsilon^{bd}
(\mathbb{Q}_{m+n-1})^{\gamma}_{\ d}\\
\ [\mathbb{B}_{m},(\mathbb{L}_{n})^{a}_{\ b} ]&=&\
[\mathbb{B}_{m},(\mathbb{R}_{n})^{\alpha}_{\ \beta} ]=\
[\mathbb{B}_{m},(\mathbb{H}_{n}) ]=0\nonumber.
\end{eqnarray}
The action of $\mathbb{B}$ on the fundamental representation
should be equal to the action of $\mathcal{T}\mathbb{H}^{-1}$.
Finally we would like to note that in the classical limit
$u\mathbb{C}= u\mathbb{C}^{\dagger}=\mathbb{H} $, just as in
\cite{Beisert:2007ty}. \smallskip

In terms of the operator $\mathbb{B}$, the proposed classical
$r$-matrix is \cite{Beisert:2007ty}
\begin{eqnarray}\label{eqn;RmatBeisert}
r_{12} &=& \frac{\mathcal{T}_{12}-\mathbb{B}\otimes
\mathbb{H}-\mathbb{H}\otimes
\mathbb{B}}{i(u_{1}-u_{2})}-\frac{\mathbb{B}\otimes
\mathbb{H} }{iu_{2}} +\frac{\mathbb{H}\otimes \mathbb{B}}{iu_{1}}+ \frac{i}{2}(u_{2}^{-1}-u_{1}^{-1})\mathbb{H}\otimes
\mathbb{H}.
\end{eqnarray}
We already know the realization of all the algebra generators on
the bound state representations, except for $\mathbb{B}$. The
operator $\mathbb{B}$ is characterized through its commutation
relations with the generators of $\mathfrak{h}$. It also coincides
with $\mathcal{T}\mathbb{H}^{-1}$ on the fundamental
representation $M=1$.
 An apparent
guess would be to identify $\mathbb{B}$  with
$\mathcal{T}\mathbb{H}^{-1}$ on the higher representations as
well. One should note, however, that this choice is not unique.
One can add to $\mathcal{T}\mathbb{H}^{-1}$ the Casimir operator
$\mathfrak{C}$ without spoiling any of the commutation relations
(\ref{Bcom}). On the fundamental representation the Casimir
vanishes and $\mathbb{B}$ coincides with
$\mathcal{T}\mathbb{H}^{-1}$. It appears that the correct
identification corresponds to taking $\mathbb{B} = \Sigma =
\mathcal{T}\mathbb{H}^{-1} + \mathfrak{C}\mathbb{H}^{-1}$.  As we
will see,  this will lead to a  complete agreement with the bound
state S-matrices in the near plane-wave limit.
\smallskip

Thus, from now on, we will be working with the following
$r$-matrix
\begin{eqnarray}\label{eqn;Rmat}
r_{12} &=& \frac{\mathcal{T}_{12}-\Sigma\otimes
\mathbb{H}-\mathbb{H}\otimes
\Sigma}{i(u_{1}-u_{2})}-\frac{\Sigma\otimes \mathbb{H} }{iu_{2}}
+\frac{\mathbb{H}\otimes \Sigma}{iu_{1}}+
\frac{i}{2}(u_{2}^{-1}-u_{1}^{-1})\mathbb{H}\otimes \mathbb{H}.
\end{eqnarray}
The last term is proportional to the identity operator and is
related to the phase factor of the S-matrix. It was shown in
\cite{Beisert:2007ty} that $r$ satisfies a number of properties
expected from a classical $r$-matrix like the classical
Yang-Baxter equation.\smallskip

Via (\ref{eqn;AlgDiff}) it is straightforward to put $r$ into
differential operator form since it is completely defined in terms
of the algebra generators and central elements. Upon taking the
near plane-wave limit, discussed below we can then compare this
operator to the S-matrix understood as a differential
operator.\smallskip

Let us give the explicit form of $r$ in terms of differential
operators and discuss some of its properties. We will consider
operators acting on $\Phi_{K}(w,\theta)\Phi_{M}(u,\vartheta)$. The
operator $\mathcal{T}_{12}$ is simple since it is composed of two
operators acting in different spaces. Writing it out is
straightforward:
\begin{eqnarray}
\mathcal{T}_{12} &=&
(-2w_{b}u_{a}+w_{a}u_{b})\frac{\partial^{2}}{\partial w_{a}\partial
u_{b}} +
(2\theta_{\beta}\vartheta_{\alpha}-\theta_{\alpha}\vartheta_{\beta})\frac{\partial^{2}}{\partial \theta_{\alpha}\partial \vartheta_{\beta}} +\nonumber\\
&& 2(\mathfrak{a}_{1}\mathfrak{d}_{2}-\mathfrak{b}_{2}\mathfrak{c}_{1})u_{a}\theta_{\alpha} \frac{\partial^{2}}{\partial w_{a}\partial \vartheta_{\alpha}} +
 2(\mathfrak{a}_{2}\mathfrak{d}_{1}-\mathfrak{b}_{1}\mathfrak{c}_{2})w_{a}\vartheta_{\alpha} \frac{\partial^{2}}{\partial u_{a}\partial \theta_{\alpha}} +\nonumber\\
&&
2(\mathfrak{a}_{2}\mathfrak{c}_{1}-\mathfrak{b}_{1}\mathfrak{d}_{2})\theta_{\alpha}\vartheta_{\beta}
\epsilon_{ab}\epsilon^{\alpha\beta}\frac{\partial^{2}}{\partial
w_{a}\partial u_{b}} +
2(\mathfrak{a}_{1}\mathfrak{c}_{2}-\mathfrak{b}_{2}\mathfrak{d}_{1})w_{a}u_{b}
\epsilon^{ab}\epsilon_{\alpha\beta}\frac{\partial^{2}}{\partial
\theta_{\alpha}\partial \vartheta_{\beta}}.
\end{eqnarray}
The coefficients
$\mathfrak{a},\mathfrak{b},\mathfrak{c},\mathfrak{d}$ are the
semi-classical limits of $a,b,c,d$ respectively. Note that the
information about the representation is completely encoded in the
coefficients $a_{i},b_{i},c_{i},d_{i}$ as well as in the action of
the differential operators on the ``short" superfields.\smallskip

Thus, the explicit form of $r$ depends quite a lot on the choice
of the bound state representations. On the other hand, the bound
state S-matrices are also quite different from each other and
hence the comparison between the two in the classical limit will
indeed be a non-trivial check of universality of the proposal.

\subsection{The near plane-wave limit}

To compare the proposed classical $r$-matrix to the bound state
S-matrices, one first has to define an appropriate limit in which
the two can be compared. This limit is called the near plane-wave
limit. The observations and analysis done here are similar to
those preformed in \cite{Torrielli:2007mc}. Let us first discuss a
suitable parameterization of $x^{\pm}_{;M}$ for a $M$-particle
bound state that allows taking the near plane-wave limit. We
identify $\hbar= g^{-1}$ and take \cite{Arutyunov:2006iu}:
\begin{eqnarray}\label{eqn;ParameterXpm}
x^{\pm}_{i;M} =
x_{i}\left(\sqrt{1-\frac{(M/g)^{2}}{(x_{i}-\frac{1}{x_{i}})^{2}}}\pm
\frac{i M/g}{x_{i}-\frac{1}{x_{i}}}\right).
\end{eqnarray}
By identifying $\hbar=g^{-1}$, it is obvious from
(\ref{eqn;ClasRmat}) that to find the classical $r$-matrix we
should expand around $g = \infty$ and work to order
$g^{-1}$.\smallskip

In this parameterization, most of the parameters simplify greatly.
For example, the central charge $H$ is given by
\begin{eqnarray}
H=M\frac{x^{2}+1}{x^{2}-1}.
\end{eqnarray}
Crossing symmetry also becomes transparent, since sending
$x_{i}^{\pm} \rightarrow \frac{1}{x_{i}^{\pm}}$ reduces to
\begin{eqnarray}
x_{i}\rightarrow \frac{1}{x_{i}}.
\end{eqnarray}
This simplifies checking crossing symmetry for the phases
encountered later on.

\subsection{The dressing phase}

The phase factors obtained from fusion and crossing symmetry are
given by (\ref{eqn;Phase}). Let us spell them out in the near
plane-wave limit since they will come into play when taking the
semi-classical limit. Consider two bound states of length
$M_{i},M_{j}$, described in the near plane-wave limit by
parameters $x_{i},x_{j}$ respectively.\smallskip

First of all, the functions $G(n)$ and the factors proportional to
the momenta are easily expanded around $g\rightarrow \infty$ by
using (\ref{eqn;ParameterXpm}):
\begin{eqnarray}
G(n) &=& 1+\frac{2 i n g^{-1}
}{x_1+\frac{1}{x_1}-x_2-\frac{1}{x_2}}+
\mathcal{O}(g^{-2})\nonumber\\
\frac{x_{j}^{+}}{x_{j}^{-}} &=& 1+2 i g^{-1} M_{j}\frac{ x_j
}{x_j^2-1}+\mathcal{O}\left(g^{-2}\right).
\end{eqnarray}
To examine the dressing phase, we first introduce the conserved
charges
\begin{eqnarray}
q_{n}(x_{i}) &=&
\frac{i}{n-1}\left(\frac{1}{(x_{i}^{+})^{n-1}}-\frac{1}{(x_{i}^{-})^{n-1}}\right)\nonumber\\
&=&2g^{-1} M_{i}\frac{x_{i}^{2-n}}{x_{i}^2-1} +
\mathcal{O}(g^{-2}).
\end{eqnarray}
The dressing phase is related to the conserved charges as follows
\begin{eqnarray}
\sigma(x_{i},x_{j}) = e^{\frac{i}{2}\theta(x_{i},x_{j})},
\end{eqnarray}
where
\begin{eqnarray}
\theta_{12} =g\sum_{r=2}^{\infty}\sum_{n=0}^{\infty} c_{r,r+1+2n}
\left(q_r\left(x_1\right)
q_{r+1+2n}\left(x_2\right)-q_r\left(x_2\right)
q_{r+1+2n}\left(x_1\right)\right),
\end{eqnarray}
with \cite{Hernandez:2006tk}
\begin{eqnarray}
c_{r,s} =
\delta_{r+1,s}-g^{-1}\frac{4}{\pi}\frac{(r-1)(s-1)}{(r+s-2)(s-r)}+\mathcal{O}(g^{-2}).
\end{eqnarray}
Since $q_{n}\sim g^{-1}$, we see that if we work to order
$g^{-1}$, it suffices to take $c_{r,s} = \delta_{r+1,s}$. Hence,
the dressing phase reduces to
\begin{eqnarray}
\theta_{12} &=&g\sum_{r=2}^{\infty}\sum_{n=0}^{\infty}
\delta_{n,0} \left(q_r\left(x_1\right)
q_{r+1+2n}\left(x_2\right)-q_r\left(x_2\right)
q_{r+1+2n}\left(x_1\right)\right)+\mathcal{O}(g^{-2})\nonumber\\
&=&g\sum_{r=2}^{\infty} \left(q_r\left(x_1\right)
q_{r+1}\left(x_2\right)-q_r\left(x_2\right)
q_{r+1}\left(x_1\right)\right)+\mathcal{O}(g^{-2})\nonumber\\
&=&4M_{i}M_{j}g^{-1}\frac{x_{i}^{2}x_{j}^{2}(x_{i}-x_{j})}{(x_{i}^2-1)(x_{j}^2-1)}\sum_{r=2}^{\infty}
\left(\frac{1}{x_{i}x_{j}}\right)^{r+1}+\mathcal{O}(g^{-2})\nonumber\\
&=&4M_{i}M_{j}g^{-1}\frac{(x_{i}-x_{j})}{(x_{i}^2-1)(x_{i}x_{j}-1)(x_{j}^2-1)}+\mathcal{O}(g^{-2}).
\end{eqnarray}
From this expression it is easy to see that, at least to first
order, the dressing phases of bound states indeed behave as stated
in \cite{Arutyunov:2008zt}.\smallskip

For example, consider a two particle bound state, described by
momenta $p_{1},p_{2}$ related by $x_{1}^{-}=x_{2}^{+}$ and a
fundamental excitation with momentum $q$. From fusion one obtains
that the total phase is given by
$\theta_{\mathrm{total}}=\theta(p_1,q)+\theta(p_2,q)$. However,
$p_{1}$ and $p_{2}$ are not independent, but since $\theta \sim
g^{-1}$ we only have to solve the condition $x_{1}^{-}=x_{2}^{+}$
up to zeroth order, which is easily seen to give
$x_{1}=x_{2}+\mathcal{O}(g^{-1})$. But this means that the phases
add and we find $\theta_{\mathrm{total}}=2\theta(p_1,q)$ to first
order, which indeed coincides with the found dressing
phase.\smallskip

To conclude, we give the total expression for the complete phase
factors (\ref{eqn;Phase}) in the near plane-wave limit:
\begin{eqnarray}
S^{AA}_{0}&=&{1+\frac{i (-1+x_{1}x_{2}) (x_1^2+x_2^2) g^{-1}
}{(-1+x_1^2) (x_1-x_2) (-1+x_2^2)}}+\mathcal{O}(g^{-2})\nonumber\\
S^{AB}_{0}&=&{1+\frac{2 i (-1+x_{1}x_{2}) (x_1^2+x_2^2) g^{-1}
}{(-1+x_1^2) (x_1-x_2) (-1+x_2^2)}}+\mathcal{O}(g^{-2})\nonumber\\
S^{BB}_{0}&=&{ 1+\frac{4 i (-1+x_1 x_2) (x_1^2+x_2^2) g^{-1}
}{(-1+x_1^2) (x_1-x_2) (-1+x_2^2)}}+\mathcal{O}(g^{-2}).
\end{eqnarray}
These phase factors will give a contribution proportional to the
identity matrix. We write
\begin{eqnarray}
\S = 1 + g^{-1} \S_{g\rightarrow\infty} + \mathcal{O}(g^{-2}).
\end{eqnarray}

\subsection{Comparison in the near plane-wave limit}

Taking the limit $g\rightarrow \infty$ for $\S^{AA},\S^{AB}$ and
$\S^{BB}$, we can compare these matrices with the proposed
universal classical $r$-matrix (\ref{eqn;Rmat}). For $\S^{AA}$
this has already been carried out in \cite{Beisert:2007ty} and
complete agreement was found. This is also the case for the
discussed bound state S-matrices.
\smallskip

Now we are ready to compare the two operators by considering their
action on all basis elements. For all the cases we find a perfect
agreement between the limiting values of the S-matrices and the
classical $r$-matrix evaluated in the corresponding bound state
representations
\begin{eqnarray}
\S_{g\rightarrow\infty}^{AA}=r^{AA}, \qquad
\S_{g\rightarrow\infty}^{AB} = r^{AB},\qquad
\S_{g\rightarrow\infty}^{BB} = r^{BB}.
\end{eqnarray}

Actually, we can do a bit more by comparing $r$ to the proposed
phase \cite{Arutyunov:2008zt} of the bound state S-matrix
$\mathbb{S}^{KM}$ corresponding to the scattering of bound states
of length $K$ and $M$. To this end, we recall that the bound state
S-matrices $\S^{KM}$ can be canonically normalized by setting the
coefficient $a_1$, which corresponds to the projector on the irrep
with maximal $\su(2)$ spin, equal to unity:
\begin{eqnarray}
\mathbb{S}_{can}^{KM} w_{1}^{K}u_{1}^{M} = w_{1}^{K}u_{1}^{M}.
\end{eqnarray}
For the fully dressed S-matrix we, therefore,  obtain
\begin{eqnarray}
\mathbb{S}^{KM} w_{1}^{K}u_{1}^{M} = S_{0}^{KM}w_{1}^{K}u_{1}^{M},
\end{eqnarray}
where $S_{0}^{KM}$ is a scalar factor given by
\cite{Arutyunov:2008zt}:
\begin{eqnarray}\label{eqn;FullPhase}
S_{0}^{KM}(x_{1},x_{2})&=&e^{\frac{a}{2} (p_{1}\epsilon_{2}-\epsilon_{1}p_{2})}\left(\frac{x_{1;K}^{-}}{x_{1;K}^{+}}\right)^{\frac{M}{2}}\left(\frac{x_{2;M}^{+}}{x_{2;M}^{-}}\right)^{\frac{K}{2}}\sigma(x_{1},x_{2})\times\nonumber\\
&&\times\sqrt{G(M-K)G(M+K)}\prod_{l=1}^{K-1}G(M-K+2l).
\end{eqnarray}
In the near plane-wave limit, this becomes:
\begin{eqnarray}
S_{0}^{KM}(x_{1},x_{2})&=&{ 1+ i KM \frac{(x_1 x_2-1)
(x_1^2+x_2^2)
 }{(x_1^2-1) (x_1-x_2) (x_2^2-1)}}g^{-1}\nonumber\\
&& -a\frac{ K M \left(x_1-x_2\right) \left(x_1
x_2-1\right)}{\left(x_1^2-1\right) \left(x_2^2-1\right)} g^{-1}
 +\mathcal{O}(g^{-2}).
\end{eqnarray}
The piece proportional to $a$ can be realized as an operator
\begin{eqnarray}
-a (u_{1}^{-1}-u_{2}^{-1}){\mathbb H}\otimes {\mathbb H}\, .
\end{eqnarray}
On the other hand, assuming that the classical $r$-matrix is
universal, we can easily compute its action on the state
$w_{1}^{K}u_{1}^{M}$. For $a=0$ we find
\begin{eqnarray}
(1+g^{-1}r) w_{1}^{K}u_{1}^{M} &=& S_{0}^{KM}(x_{1},x_{2})
w_{1}^{K}u_{1}^{M}.
\end{eqnarray}
This means that the phase factor (\ref{eqn;FullPhase}) derived in
\cite{Arutyunov:2008zt} is indeed compatible with $r$. With our
choice of $\mathbb{B}=\Sigma$, the proposed $r$-matrix
\cite{Beisert:2007ty} exhibits perfect "universality" in the sense
that it is capable of reproducing the semiclassical limit of the
quantum bound state S-matrices
$\mathbb{S}^{AA},\mathbb{S}^{AB},\mathbb{S}^{BB}$. In particular,
it correctly reproduces  the semi-classical limit of the quantum
phase $S_{0}^{KM}$ obtained from the fusion procedure. \smallskip

A last observation is that the form of the $r$-matrix is quite
simple and contains at most three derivatives, whereas an
arbitrary S-matrix $\S^{MN}$ of $M,N$ bound states would be build
up out of more complicated expressions containing higher
order differential operators. This leads to the idea that one could
use the proposed $r$-matrix to identify the non-trivial components
of the matrices $\S^{MN}$ and hopefully gain new insights in their
structure.

\section{Conclusions}

We have shown that the recently found bound state S-matrices
\cite{Arutyunov:2008zt}, $\S^{AB}$ and $\S^{BB}$ are invariant
under Yangian symmetry. In particular, Yangian invariance fixes
$\S^{BB}$ completely without appealing to the Yang-Baxter
equation.\smallskip

We have also compared the bound state S-matrices in the near
plane-wave limit to the proposed universal classical $r$-matrix of
\cite{Beisert:2007ty}. We found perfect agreement. It would be
also interesting to carry out an analogous investigation for the
$r$-matrix proposed in \cite{Moriyama:2007jt}.

\section*{Acknowledgements}

I am indebted to G. Arutyunov and S. Frolov for many valuable
discussions and to B. Eden for comments on the manuscript. Also, I
would like to thank N. Beisert for drawing my attention to the
question of how  ${\mathbb B}$ should act in the bound state
representations. This work was supported in part by the EU-RTN
network \textit{Constituents, Fundamental Forces and Symmetries of
the Universe} (MRTN-CT-2004-005104), by the INTAS contract
03-51-6346 and by the NWO grant 047017015.

\bibliographystyle{JHEP}
\bibliography{LitRmat}

\providecommand{\href}[2]{#2}\begingroup\raggedright\begin{thebibliography}{10}

\bibitem{Maldacena:1997re}
J.~M. Maldacena, {\it {The large N limit of superconformal field theories and
  supergravity}},  {\em Adv. Theor. Math. Phys.} {\bf 2} (1998) 231--252,
  [\href{http://xxx.lanl.gov/abs/hep-th/9711200}{{\tt hep-th/9711200}}].

\bibitem{Minahan:2002ve}
J.~A. Minahan and K.~Zarembo, {\it The \textrm{B}ethe-ansatz for $\mathcal{N} =
  4$ super \textrm{Y}ang-\textrm{M}ills},  {\em JHEP} {\bf 03} (2003) 013,
  [\href{http://xxx.lanl.gov/abs/hep-th/0212208}{{\tt hep-th/0212208}}].

\bibitem{Bena:2003wd}
I.~Bena, J.~Polchinski, and R.~Roiban, {\it Hidden symmetries of the
  $\mathit{AdS}_{5}\times \mathit{S}^5$ superstring},  {\em Phys. Rev.} {\bf
  D69} (2004) 046002, [\href{http://xxx.lanl.gov/abs/hep-th/0305116}{{\tt
  hep-th/0305116}}].

\bibitem{Beisert:2004hm}
N.~Beisert, V.~Dippel, and M.~Staudacher, {\it A novel long range spin chain
  and planar $\mathcal{N} = 4$ super \textrm{Y}ang- \textrm{M}ills},  {\em
  JHEP} {\bf 07} (2004) 075,
  [\href{http://xxx.lanl.gov/abs/hep-th/0405001}{{\tt hep-th/0405001}}].

\bibitem{Beisert:2005tm}
N.~Beisert, {\it {The $su(2|2)$ dynamic S-matrix}},  {\em Adv. Theor. Math.
  Phys.} {\bf 12} (2008) 945,
  [\href{http://xxx.lanl.gov/abs/hep-th/0511082}{{\tt hep-th/0511082}}].

\bibitem{Arutyunov:2006ak}
G.~Arutyunov, S.~Frolov, J.~Plefka, and M.~Zamaklar, {\it The off-shell
  symmetry algebra of the light-cone $\mathit{AdS}_{5}\times \mathit{S}^5$
  superstring},  {\em J. Phys.} {\bf A40} (2007) 3583--3606,
  [\href{http://xxx.lanl.gov/abs/hep-th/0609157}{{\tt hep-th/0609157}}].

\bibitem{Arutyunov:2004yx}
G.~Arutyunov and S.~Frolov, {\it Integrable hamiltonian for classical strings
  on $\mathit{AdS}_{5}\times \mathit{S}^5$},  {\em JHEP} {\bf 02} (2005) 059,
  [\href{http://xxx.lanl.gov/abs/hep-th/0411089}{{\tt hep-th/0411089}}].

\bibitem{Frolov:2006cc}
S.~Frolov, J.~Plefka, and M.~Zamaklar, {\it {The $\mathit{AdS}_{5}\times
  \mathit{S}^5$ superstring in light-cone gauge and its Bethe equations}},
  {\em J. Phys.} {\bf A39} (2006) 13037--13082,
  [\href{http://xxx.lanl.gov/abs/hep-th/0603008}{{\tt hep-th/0603008}}].

\bibitem{Arutyunov:2006yd}
G.~Arutyunov, S.~Frolov, and M.~Zamaklar, {\it {The Zamolodchikov-Faddeev
  algebra for $\mathit{AdS}_{5}\times \mathit{S}^5$ superstring}},  {\em JHEP}
  {\bf 04} (2007) 002, [\href{http://xxx.lanl.gov/abs/hep-th/0612229}{{\tt
  hep-th/0612229}}].

\bibitem{Serban:2004jf}
D.~Serban and M.~Staudacher, {\it {Planar N = 4 gauge theory and the Inozemtsev
  long range spin chain}},  {\em JHEP} {\bf 06} (2004) 001,
  [\href{http://xxx.lanl.gov/abs/hep-th/0401057}{{\tt hep-th/0401057}}].

\bibitem{Staudacher:2004tk}
M.~Staudacher, {\it The factorized $\mathit{S}$-matrix of \textrm{CFT/AdS}},
  {\em JHEP} {\bf 05} (2005) 054,
  [\href{http://xxx.lanl.gov/abs/hep-th/0412188}{{\tt hep-th/0412188}}].

\bibitem{Beisert:2005fw}
N.~Beisert and M.~Staudacher, {\it Long-range $\mathfrak{psu}(2,2|4)$
  \textrm{B}ethe ansaetze for gauge theory and strings},  {\em Nucl. Phys.}
  {\bf B727} (2005) 1--62, [\href{http://xxx.lanl.gov/abs/hep-th/0504190}{{\tt
  hep-th/0504190}}].

\bibitem{Kazakov:2004qf}
V.~A. Kazakov, A.~Marshakov, J.~A. Minahan, and K.~Zarembo, {\it Classical /
  quantum integrability in $\mathit{AdS/CFT}$},  {\em JHEP} {\bf 05} (2004)
  024, [\href{http://xxx.lanl.gov/abs/hep-th/0402207}{{\tt hep-th/0402207}}].

\bibitem{Arutyunov:2004vx}
G.~Arutyunov, S.~Frolov, and M.~Staudacher, {\it Bethe ansatz for quantum
  strings},  {\em JHEP} {\bf 10} (2004) 016,
  [\href{http://xxx.lanl.gov/abs/hep-th/0406256}{{\tt hep-th/0406256}}].

\bibitem{Beisert:2006ib}
N.~Beisert, R.~Hernandez, and E.~Lopez, {\it A crossing-symmetric phase for
  $\mathit{AdS}_{5}\times \mathit{S}^5$ strings},  {\em JHEP} {\bf 11} (2006)
  070, [\href{http://xxx.lanl.gov/abs/hep-th/0609044}{{\tt hep-th/0609044}}].

\bibitem{Freyhult:2006vr}
L.~Freyhult and C.~Kristjansen, {\it A universality test of the quantum string
  bethe ansatz},  {\em Phys. Lett.} {\bf B638} (2006) 258--264,
  [\href{http://xxx.lanl.gov/abs/hep-th/0604069}{{\tt hep-th/0604069}}].

\bibitem{Eden:2006rx}
B.~Eden and M.~Staudacher, {\it Integrability and transcendentality},  {\em J.
  Stat. Mech.} {\bf 0611} (2006) P014,
  [\href{http://xxx.lanl.gov/abs/hep-th/0603157}{{\tt hep-th/0603157}}].

\bibitem{Beisert:2007hz}
N.~Beisert, T.~McLoughlin, and R.~Roiban, {\it {The Four-Loop Dressing Phase of
  N=4 SYM}},  {\em Phys. Rev.} {\bf D76} (2007) 046002,
  [\href{http://xxx.lanl.gov/abs/0705.0321}{{\tt arXiv:0705.0321}}].

\bibitem{Janik:2006dc}
R.~A. Janik, {\it The $\mathit{AdS}_{5}\times \mathit{S}^5$ superstring
  worldsheet $\mathit{S}$-matrix and crossing symmetry},  {\em Phys. Rev.} {\bf
  D73} (2006) 086006, [\href{http://xxx.lanl.gov/abs/hep-th/0603038}{{\tt
  hep-th/0603038}}].

\bibitem{Beisert:2006ez}
N.~Beisert, B.~Eden, and M.~Staudacher, {\it Transcendentality and crossing},
  {\em J. Stat. Mech.} {\bf 0701} (2007) P021,
  [\href{http://xxx.lanl.gov/abs/hep-th/0610251}{{\tt hep-th/0610251}}].

\bibitem{Martins:2007hb}
M.~J. Martins and C.~S. Melo, {\it {The Bethe ansatz approach for factorizable
  centrally extended S-matrices}},  {\em Nucl. Phys.} {\bf B785} (2007)
  246--262, [\href{http://xxx.lanl.gov/abs/hep-th/0703086}{{\tt
  hep-th/0703086}}].

\bibitem{Leeuw:2007uf}
M.~de~Leeuw, {\it {Coordinate Bethe Ansatz for the String S-Matrix}},  {\em J.
  Phys.} {\bf A40} (2007) 14413--14432,
  [\href{http://xxx.lanl.gov/abs/0705.2369}{{\tt 0705.2369}}].

\bibitem{Hofman:2006xt}
D.~M. Hofman and J.~M. Maldacena, {\it {Giant magnons}},  {\em J. Phys.} {\bf
  A39} (2006) 13095--13118, [\href{http://xxx.lanl.gov/abs/hep-th/0604135}{{\tt
  hep-th/0604135}}].

\bibitem{Arutyunov:2006gs}
G.~Arutyunov, S.~Frolov, and M.~Zamaklar, {\it {Finite-size effects from giant
  magnons}},  {\em Nucl. Phys.} {\bf B778} (2007) 1--35,
  [\href{http://xxx.lanl.gov/abs/hep-th/0606126}{{\tt hep-th/0606126}}].

\bibitem{Minahan:2008re}
J.~A. Minahan and O.~Ohlsson~Sax, {\it {Finite size effects for giant magnons
  on physical strings}},  {\em Nucl. Phys.} {\bf B801} (2008) 97--117,
  [\href{http://xxx.lanl.gov/abs/0801.2064}{{\tt arXiv:0801.2064}}].

\bibitem{Klose:2008rx}
T.~Klose and T.~McLoughlin, {\it {Interacting finite-size magnons}},  {\em J.
  Phys.} {\bf A41} (2008) 285401,
  [\href{http://xxx.lanl.gov/abs/0803.2324}{{\tt arXiv:0803.2324}}].

\bibitem{Hatsuda:2008gd}
Y.~Hatsuda and R.~Suzuki, {\it {Finite-Size Effects for Dyonic Giant Magnons}},
   {\em Nucl. Phys.} {\bf B800} (2008) 349--383,
  [\href{http://xxx.lanl.gov/abs/0801.0747}{{\tt arXiv:0801.0747}}].

\bibitem{Janik:2007wt}
R.~A. Janik and T.~Lukowski, {\it {Wrapping interactions at strong coupling --
  the giant magnon}},  {\em Phys. Rev.} {\bf D76} (2007) 126008,
  [\href{http://xxx.lanl.gov/abs/0708.2208}{{\tt 0708.2208}}].

\bibitem{Gromov:2008ie}
N.~Gromov, S.~Schafer-Nameki, and P.~Vieira, {\it {Quantum Wrapped Giant
  Magnon}},  {\em Phys. Rev.} {\bf D78} (2008) 026006,
  [\href{http://xxx.lanl.gov/abs/0801.3671}{{\tt arXiv:0801.3671}}].

\bibitem{Heller:2008at}
M.~P. Heller, R.~A. Janik, and T.~Lukowski, {\it {A new derivation of Luscher
  F-term and fluctuations around the giant magnon}},  {\em JHEP} {\bf 06}
  (2008) 036, [\href{http://xxx.lanl.gov/abs/0801.4463}{{\tt
  arXiv:0801.4463}}].

\bibitem{Ambjorn:2005wa}
J.~Ambjorn, R.~A. Janik, and C.~Kristjansen, {\it {Wrapping interactions and a
  new source of corrections to the spin-chain / string duality}},  {\em Nucl.
  Phys.} {\bf B736} (2006) 288--301,
  [\href{http://xxx.lanl.gov/abs/hep-th/0510171}{{\tt hep-th/0510171}}].

\bibitem{Arutyunov:2007tc}
G.~Arutyunov and S.~Frolov, {\it {On String S-matrix, Bound States and TBA}},
  {\em JHEP} {\bf 12} (2007) 024,
  [\href{http://xxx.lanl.gov/abs/0710.1568}{{\tt 0710.1568}}].

\bibitem{Zamolodchikov:1989cf}
A.~B. Zamolodchikov, {\it {Thermodynamic Bethe Ansatz in Relativistic Models.
  Scaling Three State Potts and Lee-Yang Models}},  {\em Nucl. Phys.} {\bf
  B342} (1990) 695--720.

\bibitem{Beisert:2007ds}
N.~Beisert, {\it {The S-Matrix of AdS/CFT and Yangian Symmetry}},  {\em PoS}
  {\bf SOLVAY} (2006) 002, [\href{http://xxx.lanl.gov/abs/0704.0400}{{\tt
  0704.0400}}].

\bibitem{Matsumoto:2007rh}
T.~Matsumoto, S.~Moriyama, and A.~Torrielli, {\it {A Secret Symmetry of the
  AdS/CFT S-matrix}},  {\em JHEP} {\bf 09} (2007) 099,
  [\href{http://xxx.lanl.gov/abs/0708.1285}{{\tt arXiv:0708.1285}}].

\bibitem{Bernard:1992ya}
D.~Bernard, {\it {An Introduction to Yangian Symmetries}},  {\em Int. J. Mod.
  Phys.} {\bf B7} (1993) 3517--3530,
  [\href{http://xxx.lanl.gov/abs/hep-th/9211133}{{\tt hep-th/9211133}}].

\bibitem{MacKay:2004tc}
N.~J. MacKay, {\it {Introduction to Yangian symmetry in integrable field
  theory}},  {\em Int. J. Mod. Phys.} {\bf A20} (2005) 7189--7218,
  [\href{http://xxx.lanl.gov/abs/hep-th/0409183}{{\tt hep-th/0409183}}].

\bibitem{Moriyama:2007jt}
S.~Moriyama and A.~Torrielli, {\it {A Yangian Double for the AdS/CFT Classical
  r-matrix}},  {\em JHEP} {\bf 06} (2007) 083,
  [\href{http://xxx.lanl.gov/abs/0706.0884}{{\tt 0706.0884}}].

\bibitem{Beisert:2007ty}
N.~Beisert and F.~Spill, {\it {The Classical r-matrix of AdS/CFT and its Lie
  Bialgebra Structure}},  {\em Commun. Math. Phys.} {\bf 285} (2009) 537--565,
  [\href{http://xxx.lanl.gov/abs/0708.1762}{{\tt arXiv:0708.1762}}].

\bibitem{Matsumoto:2008ww}
T.~Matsumoto and S.~Moriyama, {\it {An Exceptional Algebraic Origin of the
  AdS/CFT Yangian Symmetry}},  {\em JHEP} {\bf 04} (2008) 022,
  [\href{http://xxx.lanl.gov/abs/0803.1212}{{\tt arXiv:0803.1212}}].

\bibitem{Dorey:2006dq}
N.~Dorey, {\it {Magnon bound states and the AdS/CFT correspondence}},  {\em J.
  Phys.} {\bf A39} (2006) 13119--13128,
  [\href{http://xxx.lanl.gov/abs/hep-th/0604175}{{\tt hep-th/0604175}}].

\bibitem{Beisert:2006qh}
N.~Beisert, {\it {The Analytic Bethe Ansatz for a Chain with Centrally Extended
  $su(2|2)$ Symmetry}},  {\em J. Stat. Mech.} {\bf 0701} (2007) P017,
  [\href{http://xxx.lanl.gov/abs/nlin/0610017}{{\tt nlin/0610017}}].

\bibitem{Chen:2006gp}
H.-Y. Chen, N.~Dorey, and K.~Okamura, {\it {The asymptotic spectrum of the N =
  4 super Yang-Mills spin chain}},  {\em JHEP} {\bf 03} (2007) 005,
  [\href{http://xxx.lanl.gov/abs/hep-th/0610295}{{\tt hep-th/0610295}}].

\bibitem{Arutyunov:2008zt}
G.~Arutyunov and S.~Frolov, {\it {The S-matrix of String Bound States}},  {\em
  Nucl. Phys.} {\bf B804} (2008) 90--143,
  [\href{http://xxx.lanl.gov/abs/0803.4323}{{\tt arXiv:0803.4323}}].

\bibitem{Gomez:2006va}
C.~Gomez and R.~Hernandez, {\it {The magnon kinematics of the AdS/CFT
  correspondence}},  {\em JHEP} {\bf 11} (2006) 021,
  [\href{http://xxx.lanl.gov/abs/hep-th/0608029}{{\tt hep-th/0608029}}].

\bibitem{Plefka:2006ze}
J.~Plefka, F.~Spill, and A.~Torrielli, {\it {On the Hopf algebra structure of
  the AdS/CFT S-matrix}},  {\em Phys. Rev.} {\bf D74} (2006) 066008,
  [\href{http://xxx.lanl.gov/abs/hep-th/0608038}{{\tt hep-th/0608038}}].

\bibitem{Kazakov:2007na}
V.~Kazakov and P.~Vieira, {\it {From Characters to Quantum (Super)Spin Chains
  via Fusion}},  \href{http://xxx.lanl.gov/abs/0711.2470}{{\tt
  arXiv:0711.2470}}.

\bibitem{Roiban:2006gs}
R.~Roiban, {\it {Magnon bound-state scattering in gauge and string theory}},
  {\em JHEP} {\bf 04} (2007) 048,
  [\href{http://xxx.lanl.gov/abs/hep-th/0608049}{{\tt hep-th/0608049}}].

\bibitem{Chen:2006gq}
H.-Y. Chen, N.~Dorey, and K.~Okamura, {\it {On the scattering of magnon
  boundstates}},  {\em JHEP} {\bf 11} (2006) 035,
  [\href{http://xxx.lanl.gov/abs/hep-th/0608047}{{\tt hep-th/0608047}}].

\bibitem{Torrielli:2007mc}
A.~Torrielli, {\it {Classical r-matrix of the $su(2|2)$ SYM spin-chain}},  {\em
  Phys. Rev.} {\bf D75} (2007) 105020,
  [\href{http://xxx.lanl.gov/abs/hep-th/0701281}{{\tt hep-th/0701281}}].

\bibitem{Arutyunov:2006iu}
G.~Arutyunov and S.~Frolov, {\it On $\mathit{AdS}_{5}\times \mathit{S}^5$
  string $\mathit{S}$-matrix},  {\em Phys. Lett.} {\bf B639} (2006) 378--382,
  [\href{http://xxx.lanl.gov/abs/hep-th/0604043}{{\tt hep-th/0604043}}].

\bibitem{Hernandez:2006tk}
R.~Hernandez and E.~Lopez, {\it Quantum corrections to the string bethe
  ansatz},  {\em JHEP} {\bf 07} (2006) 004,
  [\href{http://xxx.lanl.gov/abs/hep-th/0603204}{{\tt hep-th/0603204}}].

\end{thebibliography}\endgroup

\end{document}